\documentclass[journal]{IEEEtran}

\usepackage{xcolor,soul,framed} %,caption

\colorlet{shadecolor}{yellow}
\usepackage[pdftex]{graphicx}
\graphicspath{{../pdf/}{../jpeg/}}
\DeclareGraphicsExtensions{.pdf,.jpeg,.png}

\usepackage[cmex10]{amsmath}
\usepackage{array}
\usepackage{mdwmath}
\usepackage{mdwtab}
\usepackage{eqparbox}
\usepackage{url}
\usepackage{algorithm}
\usepackage{algorithmic}
\usepackage{caption}
\usepackage{graphicx}        
\usepackage{subcaption}

\begin{document}
\bstctlcite{IEEEexample:BSTcontrol}
    \title{Dynamic Authentication and Granularized Authorization with a Cross-Domain Zero Trust Architecture for Federated Learning in Large-Scale IoT Networks}
  \author{Xiaoyu Ma,
      Fang Fang,~\IEEEmembership{Senior Member,~IEEE,}\\
      Xianbin Wang,~\IEEEmembership{Fellow,~IEEE}% <-this % stops a space

  \thanks{Xiaoyu Ma and Xianbin Wang are with the Department of Electrical and Computer Engineering, Western University, London, ON N6A 5B9, Canada (e-mail: xma272@uwo.ca; xianbin.wang@uwo.ca).}% <-this % stops a space
  \thanks{Fang Fang is with the Department of Electrical and Computer Engineering and the Department of Computer Science, Western University, London, ON N6A 5B9, Canada (e-mail: fang.fang@uwo.ca).}}


% ====================================================================
\maketitle

% === ABSTRACT ====================================================================
% =================================================================================
\begin{abstract}
%\boldmath
With the increasing number of connected devices and complex networks involved, current domain-specific security techniques become inadequate for diverse large-scale Internet of Things (IoT) systems applications. While cross-domain authentication and authorization brings lots of security improvement, it creates new challenges of efficiency and security. Zero trust architecture (ZTA), an emerging network security architecture, offers a more granular and robust security environment for IoT systems. However, extensive cross-domain data exchange in ZTA can cause reduced authentication and authorization efficiency and data privacy concerns. Therefore, in this paper, we propose a dynamic authentication and granularized authorization scheme based on ZTA integrated with decentralized federated learning (DFL) for cross-domain IoT networks. Specifically, device requests in the cross-domain process are continuously monitored and evaluated, and only necessary access permissions are granted. To protect user data privacy and reduce latency, we integrate DFL with ZTA to securely and efficiently share device data across different domains. Particularly, the DFL model is compressed to reduce the network transmission load. Meanwhile, a dynamic adaptive weight adjustment mechanism is proposed to enable the DFL model to adapt to data characteristics from different domains. We analyze the performance of the proposed scheme in terms of security proof, including confidentiality, integrity and availability. Simulation results demonstrate the superior performance of the proposed scheme in terms of lower latency and higher throughput compared to other existing representative schemes. 
\end{abstract}

% === KEYWORDS ====================================================================
% =================================================================================

% \fi
\IEEEpeerreviewmaketitle

% ====================================================================
% ====================================================================
% ====================================================================

% === I. INTRODUCTION =============================================================
% =================================================================================
\section{Introduction}

\IEEEPARstart{T}{he} rapid evolution of Internet of Things (IoT) technologies has reshaped diverse vertical industries by enabling ubiquitous connectivity and data-aided industry automation. However, this development also introduces substantial security challenges in large scale IoT systems. Specifically, the much-needed cross-domain authentication and authorization in such IoT environments is complicated by device/network heterogeneity, diverse security requirements, and distributed resources \cite{zahra2021cross,singh2023study,istiaque2021machine,cherbal2024security}.

Due to the distributed and decentralized nature of IoT systems, security provisioning particularly authentication has to be achieved across different domains due to the involvement of different networks or subsystems with distinct security mechanisms. Current cross-domain authentication schemes in IoT system typically rely on one-time authentication \cite{liu2023blockchain,gong2023lcdma,zhang2022efficient,huang2022blockchain,jia2020irba}, where users or devices do not need to reauthenticate for each access request after initial authentication. However, this can lead to severe security risks like lateral movement attacks and session hijacking. One possible solution is to use blockchain, and securely store identity information and authentication credentials of all cross-domain devices. By utilizing smart contracts, blockchain ensures standardized and secure cross-domain identity verification \cite{zhang2022efficient,huang2022blockchain,jia2020irba,feng2021blockchain,ali2020xdbauth,zhang2024lightweight}. Despite its advantages, blockchain still faces several major difficulties in IoT applications, particularly performance issues due to transaction delays \cite{feng2021blockchain}, data privacy concerns, high energy consumption, and compatibility issues with smart contracts across different domains.

To address aforementioned challenges, Zero Trust Architecture (ZTA) has emerged as a promising solution. ZTA relies on frequently repeated authentication, which increase the opportunities to prevent potential threats. ZTA operates on the principle of \textit{never trust, always verify}, continuously monitoring and validating devices and requests to enhance security \cite{stafford2020zero}. Additionally, access control in ZTA is fine-grained, granting users and devices only the necessary permissions, helping prevent lateral movement attacks, reducing the risks of data leakage and misuse. Moreover, ZTA exhibits strong adaptability and flexibility, allowing dynamic adjustments based on changes in users, devices, and environments. However, majority of the current research on ZTA is limited to single domain security \cite{chen2023zero,mehraj2020establishing,tang2023privacy,sedjelmaci2023enabling,sedjelmaci2023zero}. In the cross-domain field, research on ZTA primarily focuses on access control design \cite{awan2023blockchain,mohseni2023real}. However, capabilities for effectively validating the identification of cross-domain devices and determining their allowed actions or privileges become crucial for the effective operation of large-scale IoT systems. Therefore, there remains a research gap in the study of cross-domain authentication and authorization by leveraging the ZTA environment. To address this, we propose a dynamic cross-domain access control and fine-grained authentication and authorization mechanism based on ZTA. This mechanism integrates continuous verification of user identities and device status with context-aware policies that adapt to factors such as device behavior. By dynamically adjusting access controls based on these contextual elements, it effectively mitigates security risks in both intra-domain and cross-domain environments.

Nonetheless, relying solely on ZTA does not adequately address cross-domain issues, as it may still face data privacy concerns, transmission efficiency problems, and increased communication overhead. Devices need to provide substantial sensitive data (e.g., location, status, access time, usage patterns) for ZTA’s scrutiny, which brings privacy leakage risks if intercepted by unauthorized parties. This could also lead to data tampering and security threats. Additionally, transmitting large amounts of ZTA related data can cause latency issues, as some mobile devices are sensitive to delays in authentication and authorization. Frequent data transmission also increases network load and resource consumption, reducing system performance and scalability.

To address the privacy and efficiency issue, in this paper, we propose integrating Decentralized Federated Learning (DFL) with ZTA to enhance collaborative learning and continuously security provisioning concurrently in cross-domain IoT environments. DFL allows for collaborative machine learning across decentralized devices while protecting data privacy, improving efficiency, and reducing latency. The operation of DFL focuses on frequently updating each node (e.g., model parameters or gradients) and metadata (e.g., activation functions in neural networks) to the remaining federated nodes \cite{beltran2023decentralized,liu2022decentralized}. When applying DFL to cross-domain authentication and authorization for ZTA, only model parameters, rather than large amounts of device information, need to be transmitted between domains. This could greatly enhance device data privacy by eliminating the need to transmit the original device information. More importantly, by transmitting the model parameters with reduced size rather than raw data, ZTA can be achieved based on DFL with significantly improved efficiency for cross-domain authorization.

In the proposed scheme, ZTA integrates with DFL to continuously and dynamically validate devices and access requests, enhancing security and efficiency in cross-domain environments. The main contributions are listed as follows:

\begin{enumerate}
    \item[1)] We propose a dynamic authentication and granularized authorization scheme for cross-domain IoT networks, integrating ZTA with DFL. The proposed scheme leverages dynamic access control and fine-grained authorization to effectively mitigate security risks in both intra-domain and cross-domain scenarios. By continuously validating access requests and device states, the proposed approach can significantly reduce the potential for unauthorized access and lateral movement attacks.
    \item[2)] To address data privacy concerns in cross-domain ZTA environments, we propose a DFL-based framework. This framework ensures that sensitive device data remains decentralized and private, thereby safeguarding user privacy. Additionally, by sharing only model parameters instead of raw data, the proposed approach can also minimize the network burden during cross-domain transmissions, thereby enhancing the efficiency of cross-domain authentication and authorization processes.
    \item[3)] To better adapt to dynamic network environments, we propose a dynamic weight adjustment mechanism within the DFL model. This mechanism allows the local model to distribute the influence of neighboring model parameters based on data characteristics and model performance across domains. This ensures that the federated learning process remains robust and accurate, maintaining high performance and adaptability based on real-time feedback.
    \item[4)] We conduct a comprehensive performance evaluation of the proposed scheme, including security proofs, simulation results, and comparisons with existing methods. The proposed scheme demonstrates lower latency and higher throughput, highlighting its effectiveness and efficiency. Through rigorous testing in simulated cross-domain authentication and authorization environments, we show that the proposed approach not only enhances security and privacy but also achieves superior performance metrics compared to traditional methods.
\end{enumerate}

The remaining sections of this paper are organized as follows. Section II discusses existing relevant research efforts, while Section III presents the intra-domain and cross-domain authentication and authorization schemes under ZTA. Furthermore, in Section IV, we propose a DFL-based scheme for pre-authorization decisions. Section V presents the security proof and simulation results of the proposed approach. Our work is summarized in Section VI.

% === II. Harmonically-Terminated Power Rectifier Analysis ========================
% =================================================================================
\section{Related Work}

Recent advancements in cross-domain authentication and authorization have focused on algorithmic improvements to enhance both security and efficiency. The core of enhancing security lies in algorithms designed to identify anomalous behaviors, which have been proven effective in reducing unauthorized access and tampering \cite{yao2015event,wang2023access}. Additionally, replacing high-complexity bilinear pair operations with symmetric polynomials has significantly reduced the computational and communication overhead for secure key exchange, thereby improving efficiency \cite{gong2023lcdma}. These innovations in algorithm design collectively contribute to more robust and efficient cross-domain authentication systems.

However, these methods tend to rely on one-time authentication and are vulnerable to lateral movement attacks and session hijacking. To enhance decentralized trust and identity-based self-authentication, some researchers have turned to blockchain \cite{liu2023blockchain,zhang2022efficient,huang2022blockchain,jia2020irba,feng2021blockchain,ali2020xdbauth}. Employing blockchain-based dynamic accumulators can effectively reduce authentication overhead, improving the efficiency of cross-domain authentication and enabling more mobile devices to participate while addressing scalability issues \cite{zhang2022efficient}. Building on this improvement, for latency-sensitive applications, integrating threshold-shared multi-signatures and smart contracts into blockchain-based 5G UAV IoT cross-domain authentication schemes provides a reliable solution \cite{feng2021blockchain}.

Despite its decentralized and secure nature, blockchain technology still faces challenges in providing fine-grained access control and achieving the necessary flexibility for various applications. To address these issues, ZTA provides a comprehensive solution \cite{chen2023zero,mehraj2020establishing,tang2023privacy,syed2022zero,liu2023secure}. It emphasizes rigorous authentication and access control, highlighting the importance of thorough device identification. The security and adaptability of ZTA have made it a suitable framework for a variety of environments \cite{syed2022zero}. In the context of 6G networks, Chen \textit{et al}. \cite{chen2023zero} proposed a software-defined ZTA, focusing on decentralized identity management and employing Third-Party Security Services (TPSS) for trust evaluation mechanisms to maintain network security. For cloud environments, Mehraj \textit{et al}. \cite{mehraj2020establishing} introduced a zero-trust strategy, addressing trust establishment challenges. In addition, in cross-domain environments, Liu \textit{et al}. \cite{liu2023secure} proposed a protocol based on sharding blockchain and partial trust under the zero-trust model, providing security guarantees for cross-domain data sharing.

To further enhance the intelligence and effectiveness of security frameworks, the integration of AI into ZTA has shown significant promise \cite{siriwardhana2021ai}. For 6G network security, integrating AI into ZTA frameworks has been proposed, utilizing adaptive algorithms and layered defense agents to bolster intrusion detection capabilities and network resilience. This includes using machine learning to identify attacks and dynamic models based on non-cooperative games, which strengthen the security of 6G edge networks \cite{sedjelmaci2023enabling,sedjelmaci2023zero}. This approach provides a robust solution for real-time threat detection and mitigation, showcasing the potential for enhanced security in next-generation networks.

Moreover, implementing zero trust based on environmental parameters and device behavior across different devices is also very crucial. Integrating Attribute-Based Access Control (ABAC) policies into IoT environments ensures context-aware access decisions. Frameworks like ZAIB (Zero-Trust and ABAC for IoT using Blockchain) leverage blockchain to enforce zero-trust principles in IoT devices \cite{awan2023blockchain}. In cloud-centric environments, real-time lightweight access control can offer fine-grained control and efficient trust assessment through attribute-based encryption and continuous trust measurements using Merkle Trees \cite{mohseni2023real}. These approaches collectively enhance authorization security by ensuring that access is continuously monitored and contextually aware.

Despite progress, current research on cross-domain authentication and authorization still faces challenges such as data privacy and efficiency. Additionally, existing studies on zero-trust authentication and authorization are limited to individual network domains, leaving a gap in research for cross-domain scenarios. To address these shortcomings, we propose a novel cross-domain authentication and authorization scheme that integrates ZTA with DFL to achieve secure access control, protect the privacy of cross-domain device data, and improve the efficiency of access authorization requests.

\section {Proposed ZTA for Intra-Domain and Cross-Domain Authentication Authorization}

This section presents the detailed architecture of the proposed system, emphasizing its components, data flow, and the intra-domain and cross-domain ZTA authentication and authorization processes. The designed architecture aims to achieve secure and fine-grained authentication and authorization in IoT networks.

\subsection{System Architecture}

\begin{figure}
  \begin{center}
  \includegraphics[width=1.0\linewidth]{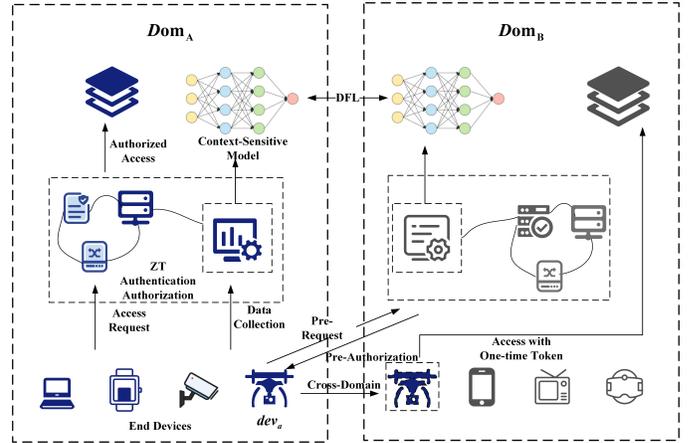}\\
  \caption{The system architecture of intra-domain and cross-domain authentication authorization under zero trust.}\label{fig_1}
  \end{center}
\end{figure}

Fig.~\ref{fig_1} illustrates the proposed architecture of the ZTA-based intra-domain and cross-domain authentication and authorization system. The architecture consists of two domains, namely $Dom_{A}$ and $Dom_{B}$\footnote{For simplicity, we use the example of two domains , which we will expand to multiple domains in later sections.}. Each domain represents a distinct administrative boundary with its own devices and policies, including mobile devices like drones, phones, and smartwatches. When a device needs to access a resource, it must be authenticated and authorized through the ZTA, ensuring only authorized requests can access the specified resource.

The authentication and authorization process of ZTA is managed by various edge devices and edge gateways. Edge devices, such as sensors, make real-time decisions at the data generation points. Edge gateways handle data aggregation, preprocessing, and secure communications. Continuous monitoring and dynamic evaluation within the ZTA framework provide fine-grained access control, ensuring secure handling of all requests in both local and cross-domain environments.

\begin{figure}
  \begin{center}
  \includegraphics[width=1.0\linewidth]{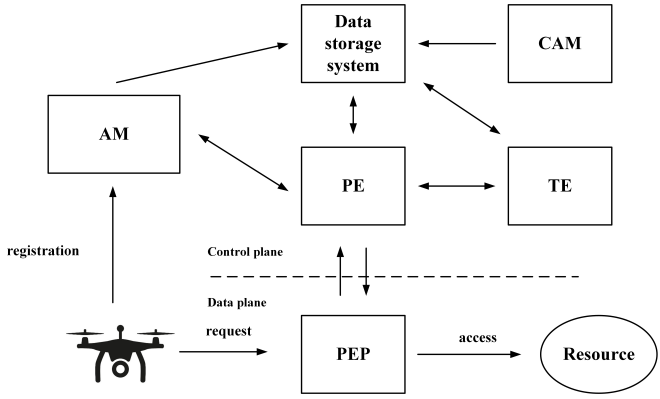}\\
  \caption{The framework of ZTA authentication authorization.}\label{fig_2}
  \end{center}
\end{figure}

Fig.~\ref{fig_2} illustrates the core components of ZTA and its integrated authentication and authorization framework. In ZTA, the data plane manages network traffic and protection, while the control plane oversees authentication and access control. The Authentication Module (AM) registers and authenticates devices, after which the Policy Enforcement Point (PEP) forwards access requests to the Policy Engine (PE) for authorization. The PE collaborates with the Trust Engine (TE) and data storage system, leveraging real-time context from the Context-Aware Module (CAM) to perform dynamic risk assessments and enforce security policies. Based on the PE's decision, the PEP either grants or denies access to resources, ensuring secure, context-aware data flows within the zero-trust environment \cite{gilman2017zero}.

Existing cross-domain authentication authorization methods usually face some security issues such as lack of dynamic risk assessment and real-time context-awareness for authentication. To address these issues, we propose a cross-domain authorization and authentication scheme based on ZTA. For cross-domain access, as shown in Fig. 1, for example, device $dev_{a}$ from $Dom_{A}$ initiates a cross-domain access request to the target domain, $Dom_{B}$. The device information from $Dom_{A}$ is passed through a secure channel to $Dom_{B}$ for evaluation and pre-authorization decision. After the decision is passed, $Dom_{B}$ issues a one-time token to $dev_{a}$. When $dev_{a}$ enters into $Dom_{B}$, it can directly access the specified resources with this token without having to go through the authentication authorization process again.

To further address privacy and efficiency issues in cross-domain data transfers, we integrate DFL into the ZTA. Each network domain maintains a device context prediction model locally and collaborates with neighboring domains through DFL for joint training. During training, only model parameters are transmitted, protecting data privacy and improving transmission efficiency. With DFL model training, each network domain is able to predict the context data of devices in neighboring domains to optimize zero-trust pre-authorization decision, which can be referred to Section IV.

\subsection{Intra-Domain Authentication Authorization}

In this section, we will introduce the procedures for intra-domain authentication and authorization.

\begin{figure}
  \begin{center}
  \includegraphics[width=0.7\linewidth]{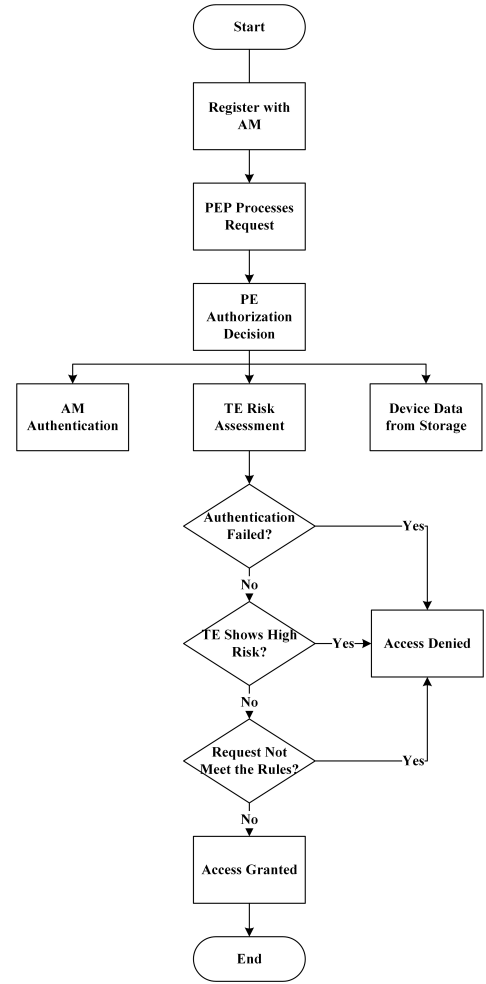}\\
  \caption{The flowchart of the authentication authorization process of the device under ZTA.}\label{fig_3}
  \end{center}
\end{figure}

Fig.~\ref{fig_3} represents a flowchart of the authentication authorization process of the device under ZTA.

\begin{enumerate}
    \item[\textit{1)}] \textit{Registration:} When $dev_{a}$ first enters $Dom_{A}$, it must register by generating a key pair $\left ( Pk_{a}, Sk_{a}  \right )$ and sending the public key and device details to the AM. The AM, assumed to be trusted, verifies the information, signs the public key with its private key $Sk_{AM}$, and issues a digital certificate $Cert_{a} $ that includes $dev_{a}$'s unique identifier $ID_{a}$. This certificate is stored in the data storage system, allowing other network entities to verify the $dev_{a}$'s authenticity. 
    \item[\textit{2)}] \textit{Issue a request:} When $dev_{a}$ requests access to $resource\_i$, it sends an access request $R_{a}^{i}$ containing $Cert_{a}$, $ID_{a}$, $resource\_i$, $access\_level$ and $access\_intention$.
    \item[\textit{3)}] \textit{Processing requests:} The PEP intercepts this request and forwards it to the PE for authorization. The PE consults the AM to authenticate $dev_{a}$ by verifying its digital certificate. If verified, the AM sends the result to the PE. The TE then assesses the risk of $R_{a}^{i}$ using contextual information from the data storage system, updated in real-time by the CAM. The TE uses static rules and machine learning for trust scoring. The PE combines information from the AM, TE, and data storage to make an authorization decision and sends the result $A_{a}^{i}$ to the PEP.
    \item[\textit{4)}] \textit{Make decisions:} If any verification step fails, or if the risk is deemed too high, or if the request does not comply with security policies, the access request is denied. Otherwise, the access is granted, ensuring secure and authorized access within the domain.
\end{enumerate}

\subsection{The Proposed Cross-Domain Authentication Authorization}

Based on the intra-domain authentication and authorization framework, we further propose a novel framework to support cross-domain authentication and authorization. As the registration process for devices remains the same as in the intra-domain scenario, it will not be described here.

By integrating ZTA with DFL, we present a novel approach to enhance cross-domain authentication and authorization while protecting privacy. This method leverages the continuous collection of contextual information within each domain to enable secure and efficient pre-authorization processes. Taking Fig. 1 as an example, the CAMs located in the $Dom_{A}$ and $Dom_{B}$ continuously collect the context information of the devices in real time. These collectively form the device context dataset, denoted as $D_{A}$ and $D_{B}$, for each respective domain. Within the CAMs, $Dom_{A}$ and $Dom_{B}$ each use their respective datasets $D_{A}$ and $D_{B}$ to train models $M_{A}$ and $M_{B}$. These models enable the prediction of detailed device context information based on basic device identifiers (such as unique device identifiers), facilitating pre-authorization. 

Additionally, $Dom_{A}$ and $Dom_{B}$ periodically share model parameters to conduct DFL. Therefore, $Dom_{A}$ and $Dom_{B}$ can use the models to obtain contextual information about devices requesting cross-domain access. Since the training of models $M_{A}$ and $M_{B}$ occurs independently within each domain, without sharing $D_{A}$ and $D_{B}$, there is no risk of leaking information about the devices. This ensures the privacy and security of device information. The detailed training process of DFL is described in Section IV.

We summarized the proposed DFL-based cross-domain pre-authorization process in Algorithm 1. When a mobile $dev_{a}$ in $Dom_{A}$ wants to move to $Dom_{B}$ for access, it needs to send an access request, denoted as $R_{a}^{j}=$($Cert_{a}$, $ID_{a}$, $Dom_{B}$, $resource\_j$, $access\_level$, $access\_intention$), to the PEP. Upon receiving the request, the PEP first checks the network area where the requested resource resides. If the requested resource is within the local domain, the PEP forwards the request $R_{a}^{j}$ to the PE for further decision-making. If the requested resource is in the neighboring domain, the PEP securely communicates the request $R_{a}^{j}$ to neighboring domain $Dom_{B}$. $Dom_{B}$ then conducts pre-authorization decision-making on the request $R_{a}^{j}$ for $dev_{a}$.

To ensure secure communication between $Dom_{A}$ and $Dom_{B}$, we apply elliptic curve cryptography (ECC) to encrypt the communication content (Line 4-6 of Algorithm 1). ECC involves selecting elliptic curve parameters, generating private and public keys for each domain, exchanging these public keys, and computing a shared key for secure communication using the elliptic curve equation. Thus it provides high security and efficiency compared to traditional methods by offering equivalent security with smaller key sizes, thereby reducing computational overhead \cite{pub2000digital,menezes1997handbook}. Based on the key $K$ generated by the ECC algorithm, the request encryption process between neighboring domains in the proposed scheme includes the following steps:

\begin{enumerate}
    \item[\textit{1)}] \textit{Serialize access request:} The PEP in $Dom_{A}$ converts the received request from $dev_{a}$, $R_{a}^{j}=$($Cert_{a}$, $ID_{a}$, $Dom_{B}$, $resource\_j$, $access\_level$, $access\_intention$), into a byte-stream format using standard serialization methods.
    \item[\textit{2)}] \textit{Encrypt using shared key:} The serialized access request $R_{a}^{j}$ is used as input data. Using the shared key $K$ as the key for a symmetric encryption algorithm, $R_{a}^{j}$ is encrypted to generate ciphertext $C_{a}^{j}$, which contains the encrypted data of $dev_{a}$'s access request for $resource\_j$ in $Dom_{B}$.
    \item[\textit{3)}] \textit{Send encrypted access request:} $Dom_{A}$ sends the encrypted access request $C_{a}^{j}$ to $Dom_{B}$.
    \item[\textit{4)}] \textit{Decrypting the cipher:} Upon receiving the ciphertext $C_{a}^{j}$, $Dom_{B}$ uses the same shared key $K$ and symmetric decryption algorithm to decrypt the ciphertext. The decryption operation restores the original access request $R_{a}^{j}$ in byte-stream format.
\end{enumerate}

Upon receiving $R_{a}^{j}$, $Dom_{B}$ inputs $dev_{a}$'s unique identifier $ID_{a}$ into the model $M_{B}$ to obtain detailed contextual information of $dev_{a}$ (refer to Section IV for details). Subsequently, the PEP in $Dom_{B}$ processes the request $R_{a}^{j}$ from $dev_{a}$ in $Dom_{A}$. Based on the request information and the device data obtained from the model, the PEP will perform ZTA authentication and authorization, including identity authentication by the AM, risk assessment by the TE, and the final authorization decision by the PE. Finally, PE sends the authorization decision $A_{a}^{j}$ to AM (Line 8-11 of Algorithm 1).

If the authorization decision $A_{a}^{j}$ allows access, the PE needs to inform AM of the scope $S_{j}$ of resources or services authorized for $dev_{a}$, the time range $T_{a}$ for access authorization, and also specify the specific purpose or behavior $I_{a}$ of $dev_{a}$ accessing the resources. AM will generate a one-time token $Token_{a}$ for $dev_{a}$ based on this information (Line 13-18 of Algorithm 1):

\begin{enumerate}
    \item[\textit{1)}] \textit{Generate random number:} Let $r$ be a securely generated random number. The length of $r$ should be long enough to ensure randomness and uniqueness.
    \item[\textit{2)}] \textit{Token content:} $Token_{a}$ includes the $dev_{a}$'s unique identifier $ID_{a}$, the scope $S_{j}$ of authorized access to resources or services, the time range $T_{a}$ of access authorization, the purpose or behavior $I_{a}$ of access, and the random number $r$.
    \item[\textit{3)}] \textit{Signing:} Use the private key $Sk_{AM}$ of AM to sign the content of the token $Token_{a}$, generating the signature $Sig_{AM}(Token_{a})$.
    \item[\textit{4)}] \textit{Send token:} AM sends the one-time token $Token_{a}$ with the signature $Sig_{AM}(Token_{a})$ to $dev_{a}$.
\end{enumerate}

\begin{algorithm}[t]
\caption{Cross-Domain Pre-Authorization.}\label{alg:alg1}
\begin{algorithmic}[1]
\STATE \textbf{Input:} Access request $R_{a}^{j}=$($Cert_{a}$, $ID_{a}$, $Dom_{B}$, $resource\_j$, $access\_level$, $access\_intention$) from $dev_{a}$ in $Dom_{A}$ to access $resource\_j$ in $Dom_{B}$
\STATE \textbf{Procedure:}
\STATE \quad \textbf{$\mathbf{Dom_{A} }$  Processing:}
\STATE \quad \quad Serialize access request $R_{a}^{j}$
\STATE \quad \quad $C_{a}^{j} \leftarrow \text{Encrypt}_{K}(\text{Serialized\_R}_{a}^{j})$
\STATE \quad \quad $C_{a}^{j} \rightarrow \text{$Dom_{B}$}$
\STATE \quad \textbf{$\mathbf{Dom_{B} }$ Processing:}
\STATE \quad \quad $R_{a}^{j} \leftarrow \text{Decrypt}_{K}(C_{a}^{j})$
\STATE \quad \quad $context\_info_{a} \leftarrow M_{B}(\text{Get info of } dev_{a} )$
\STATE \quad \quad $A_{a}^{j} \leftarrow PE(\text{Handle } R_{a}^{j})$
\STATE \quad \quad $A_{a}^{j} \rightarrow AM$
\STATE \quad \quad \textbf{Generate one-time token:}
\STATE \quad \quad \textbf{If} $A_{a}^{j}$ allows access \textbf{then}
\STATE \quad \quad \quad $S_j, T_a, I_a \leftarrow \text{Specify scope, time range, and intention}$
\STATE \quad \quad \quad Generate random number $r$
\STATE \quad \quad \quad $Token_{a} \leftarrow (ID_a, S_j, T_a, I_a, r)$
\STATE \quad \quad \quad $Sig_{AM}(Token_{a}) \leftarrow \text{Sign}_{Sk_{AM}}(Token_{a})$
\STATE \quad \quad \quad $(Token_{a}, Sig_{AM}(Token_{a}))\rightarrow dev_{a} $
\STATE \textbf{Output:} One-time token $Token_{a}$, $Sig_{AM}(Token_{a})$
\end{algorithmic}
\label{alg1}
\end{algorithm}

Algorithm 2 outlines the process for verifying a one-time token when a device requests access to the target domain. When $dev_{a}$ receives a one-time token $Token_{a}$ from $Dom_{B}$, it needs to enter $Dom_{B}$ and access services within the time range $T_{a}$ specified by the token. Upon entering $Dom_{B}$, $dev_{a}$ sends an access request $R_{a}^{j}=$($Cert_{a}$, $ID_{a}$, $Dom_{B}$, $resource\_j$, $access\_level$, $access\_intention$) to the PEP, along with the one-time token $Token_{a}$ carrying the signature $Sig_{AM}(Token_{a})$. Upon receiving $Token_{a}$, PEP needs to verify it :

\begin{enumerate}
    \item[\textit{1)}] \textit{Verify signature:} PEP sends the signature $Sig_{AM}(Token_{a})$ to AM for verification. AM uses the public key $Pk_{AM}$ to verify it, ensuring the legitimacy of the token.
    \item[\textit{2)}] \textit{Check authorization and time range:} Verify if the resource scope $S_{j}$ and the access authorization time range $T_{a}$ in the token match the current request.
    \item[\textit{3)}] \textit{Verify access purpose or behavior:} Check if the access purpose or behavior $I_{a}$ in the token matches the current request.
    \item[\textit{4)}] \textit{Mark token as used:} Once the token is verified and access is authorized, mark the token as used.
\end{enumerate}

When authentication passes, $dev_{a}$ is authorized to access the resources it requested.

\begin{algorithm}[t]
\caption{Real-Time Authentication Authorization for Cross-Domain Devices.}\label{alg:alg2}
\begin{algorithmic}[2]
\STATE \textbf{Input:} $Token_{a}$, $Sig_{AM}(Token_{a})$
\STATE \textbf{Procedure:}
\STATE \quad \quad Verify $Sig_{AM}(Token_{a})$ with $Pk_{AM}$
\STATE \quad \quad Check $t \in T_a$
\STATE \quad \quad Check $resource\_j \in S_j$
\STATE \quad \quad Verify $I_{a}$
\STATE \quad \quad Check if $Token_a$ is unused
\STATE \quad \quad \textbf{If} all above verifications pass \textbf{then}
\STATE \quad \quad \quad Mark $Token_{a}$ as used
\STATE \quad \quad \quad Grant access to $resource_j$
\STATE \quad \quad \textbf{else}
\STATE \quad \quad \quad Deny access
\end{algorithmic}
\label{alg2}
\end{algorithm}

\section{DFL for Cross-Domain Pre-Authorization}
In Section III, we propose a framework that enables ZTA fine-grained pre-authorization through shared device information between domains. To further achieve secure and efficient cross-domain authentication and authorization, in this section, we will address the issue of sharing device context information in cross-domain pre-authorization with the help of DFL.

\subsection{Method Implementation}

In cross-domain IoT environments, ensuring secure and efficient device authorization is crucial. Traditional methods, such as centralized authorization systems, often face challenges such as high latency, data privacy concerns, and increased network load. To address these issues, we propose integrating DFL within the ZTA framework. DFL enables collaborative learning across multiple domains without sharing raw data, thereby enhancing privacy and reducing transmission load. 

In the proposed framework, DFL models are trained locally within each domain by using both local device data and model parameters from neighboring domains to create a more comprehensive and accurate model across all domains. This approach enables each domain to predict the context data of neighboring devices, facilitating better ZTA pre-authorization decisions. Only model parameters, not raw device data, are shared between domains. Each domain can access context information solely based on the unique identifier of devices sending cross-domain requests, ensuring secure and legality data usage. The updated model aids the AM and PE in making informed pre-authorization decisions for cross-domain devices.

\subsection{DFL Model Construction}

\begin{figure*}
  \begin{center}
  \includegraphics[width=0.8\linewidth]{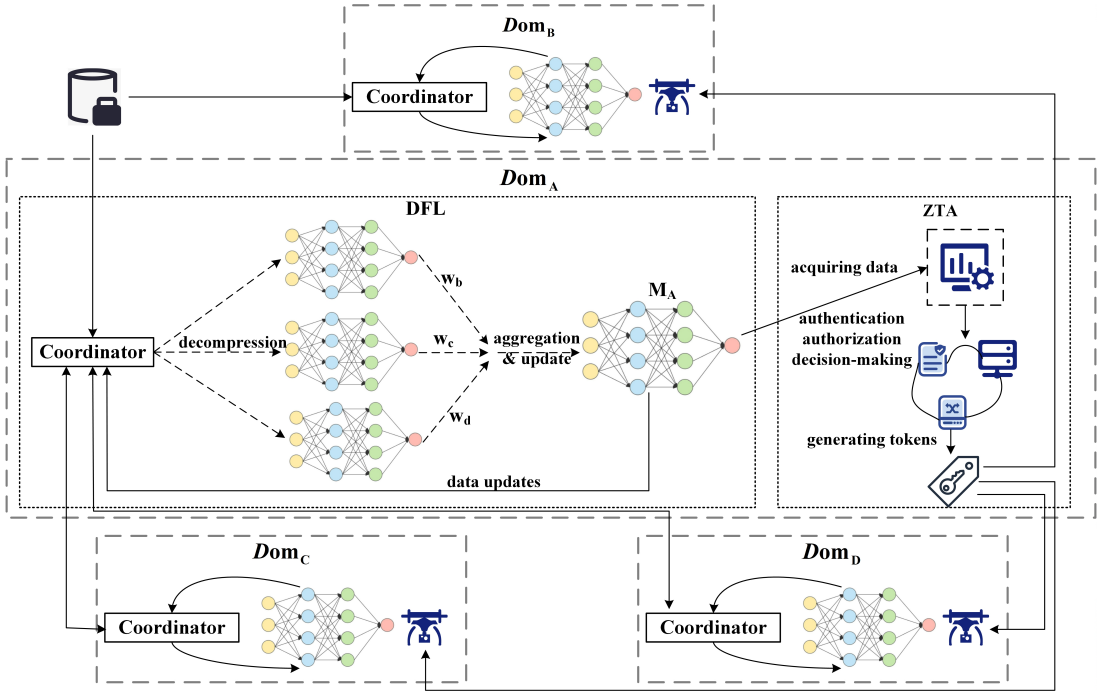}\\
  \caption{Integration of DFL into ZTA.}\label{fig_4}
  \end{center}
\end{figure*}

Fig.~\ref{fig_4} demonstrates the proposed DFL based ZTA. We set up a worker in each domain, which collaborates to train DFL models on the device datasets of each domain. These distributed workers are edge nodes on each network region, denoted as $N=\left \{ 1,2,...,n \right \}$. Unlike traditional DFL \cite{wang2022accelerating,roy2019braintorrent}, in our research context, mobile devices only move between adjacent network domains. Therefore, each worker only needs to exchange model parameters with its neighboring workers. As shown in Fig. 4, $Dom_{B}$, $Dom_{C}$, and $Dom_{D}$ are all neighboring domains of $Dom_{A}$. As each worker is located in a different network domain, the types of devices contained in different network domains may vary significantly. Consequently, the device contextual information collected by different network domains may also differ substantially. Therefore, we default that the device context datasets trained by each worker are non-IID, denoted as $D_{i}$, $i=1,2,...,n$.

Each worker maintains a deep learning model $M_{i}$ within its respective network domain. Each model is trained based on the local device contextual  dataset $D_{i}$ and neighboring model parameters. Therefore, each model can acquire device context information based on unique identifiers from the local or neighboring domains, enabling further pre-authorization decision-making. DFL iteratively trains each model until convergence. Each network domain has a federated learning coordinator $C_{i}$, responsible for managing local data, coordinating model parameters and aggregation, as well as monitoring and adjusting the federated learning process.

To reduce the amount of data transmitted in each round of DFL training, we employ model compression techniques on the local models in each domain. Each round involves transmitting only the compressed models between adjacent network domains. Additionally, we propose a dynamic weight adjustment mechanism for the models. Based on feedback mechanisms, the model weights are dynamically adjusted according to the performance of each network domain's model and the quality of local data. This enables the training models to better adapt to the data characteristics of different neighboring domains.

\subsection{Model Compression}

In the DFL training process for cross-domain pre-authorization, we exploit model compression techniques to enhance efficiency by reducing data transfer between domains. The TopK algorithm \cite{liu2015sparse} retains only the K parameters with the largest absolute values, minimizing model size while preserving key parameters that impact performance. This approach optimizes the balance between model performance and efficiency.

In the proposed scheme, during the $t$-th round of training in DFL, the local model for worker $i$ is denoted as $\mathbf{m}_{i}^{\left ( t \right )}$, consists of weight vectors $\mathbf{w}_{ij}$ for each of the $l$ layers, i.e., $\mathbf{m}_{i}^{\left ( t \right )}=\left \{ \mathbf{w}_{i1}^{(t)},\mathbf{w}_{i2}^{(t)},...,\mathbf{w}_{il}^{(t)} \right \}$. After the $t$-th round of training, the local model $\mathbf{m}_{i}^{\left ( t \right )}$ is compressed into a sparse vector $\hat{\mathbf{m}}_{i}^{\left ( t \right )}$, where each weight parameter vector is compressed into a sparse form.

Specifically, for each weight parameter $\mathbf{w}_{ij}^{(t)}$, we calculate its absolute value and select the top-K with the largest values, forming an index set $l$. Parameters outside this set are set to zero, resulting in a sparse vector $\hat{\mathbf{m}}_{i}^{\left ( t \right )}$. Therefore, each parameter $\mathbf{w}_{ij}^{(t)}$ is:

\begin{equation}
\hat{\mathbf{w}}_{ij} = \begin{cases}
\mathbf{w}_{ij}^{(t)}, & \text{if } j \in I \\
0, & \text{otherwise.}
\end{cases}
\end{equation}

\noindent where $\hat{\mathbf{w}}_{ij}^{(t)}$ represents the $j$-th parameter of the sparse vector $\hat{\mathbf{m}}_{i}^{\left ( t \right )}$.

Through this process, the local model of worker $i$ is compressed into a sparse vector after each round of training, facilitating model transmission and aggregation.

\subsection{The Proposed Dynamic Weight Adjustment}

Due to the potentially varying device data distribution and data quality across different network domains, we propose to dynamically adjust weight parameters to adaptively account for the unique characteristics of each network domain. Dynamic weight adjustment is important in multi-domain environments because it allows the model to effectively capture domain-specific patterns and variations, and thus improve the overall accuracy and robustness of the model.

To achieve dynamic weight adjustment in a DFL environment, we need to evaluate the performance of each domain's model and the relevance of its data. To this end, we have designed a feedback mechanism based on F1 score and Kullback-Leibler (KL) divergence. The F1 score, as the main index for evaluating the performance of the model, integrates the precision and recall of the model, and can evaluate the classification effect of the model in a more comprehensive way. F1 score can be calculated as:

\begin{equation}
F1=\frac{Precision\times Recall}{Precision+Recall}
\end{equation}

\noindent where $Precision$ represents the proportion of samples classified as positive by the classifier that are truly positive; $Recall$ represents the proportion of positive samples that are correctly classified as positive by the classifier.

The KL divergence is employed as the metric to measure the difference in data distribution between different network domains. Assuming that the discrete probability distributions of device data in $Dom_{A}$ and $Dom_{B}$ are denoted as $P_{A}$ and $P_{B}$ respectively, the formula for computing the KL divergence, which measures the difference between these two probability distributions, is as follows:

\begin{equation}
D_{KL}(P_{A}||P_{B})=\sum_{i}P_{A}(i)log(\frac{P_{A}(i)}{P_{B}(i)})
\end{equation}

\noindent where, $P_{A}\left ( i \right )$ and $P_{B}\left ( i \right )$ respectively represent the probabilities of obtaining device context information $i$ under probability distributions $P_{A}$ and $P_{B}$.

Note that KL divergence is not a symmetric measure, i.e., $D_{KL}(P_{A}||P_{B}) \neq D_{KL}(P_{B}||P_{A})$. Therefore, when the network domain with probability distribution $P_{B}$, $Dom_{B}$, receives the probability distribution $P_{A}$ from $Dom_{A}$, the calculation should be $D_{KL}(P_{A}||P_{B})$, indicating the measure of $P_{A}$ with respect to $P_{B}$.

Then, we calculate the weight adjustment factor based on the F1 score and KL divergence:

\begin{equation}
\textit{waf} = \lambda_{1} \textit{F1 score} + \lambda_{2} \textit{KL Divergence}
\end{equation}

\noindent where $\lambda_{1}$ and $\lambda_{2}$ denote the weight coefficients of the F1 score and KL divergence, respectively.

In the proposed scheme, the weight adjustment factor $waf$ reflects the contribution level of the domain during the training process. Therefore, in each round of training, the coordinator dynamically adjusts the model parameters and learning rates of local models based on weight adjustment factors from all neighboring domains to accommodate changes in domain contributions.

\subsection{Training Process}

In the DFL training process, each domain collaborates with its neighboring domains to train models, aiming to securely obtain information from neighboring devices and make pre-authorization decisions for incoming neighboring devices. We provide a description of the training process for worker and coordinator in each domain in Algorithm 3.

\begin{algorithm}[t]
\caption{Training Procedure of Domain $i$.}\label{alg:alg3}
\begin{algorithmic}[3]
\STATE \textbf{Initialization:} Set $\eta_{0}$, $waf_{d}^{(0)}$ ,$\alpha_{d}^{(0)}$ for each neighboring domain
\FOR{$t=1$ \textbf{to} $T$}
\STATE \quad Exchange $\hat{m}^{(t)}$ (F1 scores and probability distributions)
\STATE \quad Decompress $\hat{m}_{d}^{(t)} \rightarrow \tilde{m}_{d}^{(t)}$
\STATE \quad Calculate $waf_{d}^{(t)}$ by Eq. (4)
\STATE \quad Normalize $waf_{d}^{(t)} \rightarrow w_{d}^{(t)}$ by Eq. (5)
\STATE \quad Aggregation model parameters
\STATE \quad Calculate $\eta_{i}^{(t)}$ by Eq. (7)-(9)
\STATE \quad Perform local updating by Eq. (10)
\STATE \quad Compress local model $\rightarrow \hat{m}_{i}^{(t+1)}$
\STATE \quad Calculate F1 score and probability distribution
\STATE \quad Update $\alpha _{d}^{(t+1)}\leftarrow \alpha _{d}^{(t)}+\beta \cdot \Delta \alpha _{d}^{(t)}$
\STATE \quad Send $\hat{m}_{i}^{(t+1)}$, F1 score, and probability distribution to neighboring domains
\ENDFOR
\end{algorithmic}
\label{alg3}
\end{algorithm}

\begin{enumerate}
    \item[\textit{1)}] \textit{Initial setup:} During the initial training phase, the coordinator of each domain needs to set a baseline learning rate $\eta_{0}$ for the model and initialize the weight adjustment factors $waf_{d}^{(0)}$ and learning rate adjustment parameters $\alpha_{d}^{(0)}$ for each neighboring domain.
    \item[\textit{2)}] \textit{Sharing parameters:} At the $t$-th round of training, we denote the model of worker $i$ as $m_{i}^{(t)}$ and the models of all neighboring workers of $i$ as $m_{d}^{(t)}$. The coordinator first sends and receives the compressed model parameters $\hat{m}_{d}^{(t)}$, F1 scores, and probability distributions $P_{d}$ from all neighboring domains. Then, the coordinator decompresses all received compressed model parameters $\hat{m}_{d}^{(t)}$ to $\tilde{m}_{d}^{(t)}$ (Line 3-4 of Algorithm 3).
    \item[\textit{3)}] \textit{Calculate the model parameters:} The coordinator computes the weight adjustment factor $waf_{d}^{(t)}$ for each domain based on the received F1 scores and probability distributions $P_{d}$ as described in Eq. (4), and then normalizes the weight adjustment factors through the softmax function to obtain the weight $w_{d}^{(t)}$ for each domain (Line 5-6 of Algorithm 3):

\begin{equation}
w_{d}^{(t)}=\frac{e^{waf_{d}^{(t)}}}{\sum_{i=1}^{n}e^{waf_{i}^{(t)}}}
\end{equation}

\noindent where, $n$ represents the total number of neighboring domains, and $\sum_{i=1}^{n}e^{waf_{i}^{(t)}}$ denotes the sum of exponentiated weight adjustment factors for all neighboring domains.

Then, the model parameters $\tilde{m}_{d}^{(t)}$ are aggregated using a weighted average to obtain the local domain model parameters (Line 7 of Algorithm 3):

\begin{equation}
\tilde{m}_{i}^{(t)}=\sum_{d=1}^{n}w_{d}^{(t)}\cdot \tilde{m}_{d}^{(t)}
\end{equation}

    \item[\textit{4)}] \textit{Calculate the learning rate:} For each neighboring domain $d$, the coordinator dynamically computes the learning rate adjustment $\Delta \eta _{d}^{(t)}$ based on the weight adjustment factor $waf_{d}^{(t)}$ and the learning rate adjustment parameter $\alpha _{d}^{(t)}$ of the domain:

\begin{equation}
\Delta \eta _{d}^{(t)}=f\left ( waf_{d}^{(t)},\alpha _{d}^{(t)} \right )
\end{equation}

\noindent where $f$ is a function that can be determined based on experimentation and experience. Then, the learning rate adjustment $\Delta \eta _{d}^{(t)}$ is applied to the baseline learning rate to obtain the actual learning rate $\eta_{d}^{(t)}$ for domain $d$:

\begin{equation}
\eta _{d}^{(t)}=\eta _{0}+\Delta \eta _{d}^{(t)}
\end{equation}

Subsequently, the coordinator aggregates the actual learning rates from all neighboring domains to obtain the overall learning rate $\eta_{i} ^{(t)}$ for the current domain's $t$-th round of training (Line 8 of Algorithm 3):

\begin{equation}
\eta_{i} ^{(t)} =\frac{1}{n}\sum_{d=1}^{n}\eta _{d}^{(t)}
\end{equation}

    \item[\textit{5)}] \textit{Update the local model:} The worker uses the aggregated model parameters $\tilde{m}_{i}^{(t)}$ and the learning rate $\eta_{i}^{(t)}$ to update its local model. The worker updates its local model via gradient descent as follows (Line 9 of Algorithm 3):

\begin{equation}
m_{i}^{(t+1)}=m_{i}^{(t)}-\eta _{i}^{(t)}\cdot \nabla J(\tilde{m}_{i}^{(t)})
\end{equation}

\noindent where $J(\tilde{m}_{i}^{(t)})$ is the gradient of the loss function $J$ with respect to the local model parameters $\tilde{m}_{i}^{(t)}$.

    \item[\textit{6)}] \textit{Update parameters:} After the local model training is completed, the worker compresses the trained model to $\hat{m}_{i}^{(t+1)}$ according to Eq. (1). At the same time, the coordinator calculates the F1 score of the model $m_{i}^{(t+1)}$ and concurrently computes the probability distribution $P_{i}$ of the current data. In addition to this, the coordinator needs to adaptively updates the learning rate adjustment parameters $\alpha_{d}$ for all neighboring domains (Line 10-12 of Algorithm 3):

\begin{equation}
\alpha _{d}^{(t+1)}=\alpha _{d}^{(t)}+\beta \cdot \Delta \alpha _{d}^{(t)}
\end{equation}

\noindent here, $\beta$ is a smoothing factor used to control the rate of change of the learning rate adjustment parameters. $\Delta \alpha _{d}^{(t)} = f(waf_{d}^{(t)}, \gamma_{d}^{(t)})$ represents the parameter adjustment amount, which is determined based on the coordinator's real-time monitoring of the neighboring domains. In this context, $waf_{d}^{(t)}$ denotes the weight adjustment factor calculated by the coordinator for neighboring domain $d$ in the $t$-th round, and $\gamma_{d}^{(t)}$ represents the performance metric received (monitored) by the coordinator from neighboring domain $d$. Here, we utilize the F1 score as its performance metric.

Finally, the coordinator sends the compressed model $\hat{m}_{i}^{(t+1)}$, F1 scores, and probability distribution $P_{i}$ to all neighboring domains to initiate a new round of training (Line 13 of Algorithm 3).
\end{enumerate}

This iterative process ensures that the model continuously improves by leveraging diverse data from multiple domains while preserving privacy and reducing communication overhead.

\section{Evaluation}
In this section, we first conduct a security analysis of the proposed cross-domain authentication and authorization scheme, followed by simulations to evaluate its performance.

\subsection{Security Analysis}

\subsubsection{Confidentiality}

The proposed scheme ensures confidentiality through the integration of DFL within the ZTA. By utilizing DFL, sensitive device data is kept within its origin domain, as only model parameters are shared with neighboring domains. This approach minimizes the risk of data exposure during cross-domain communications, as no raw data is transmitted. Additionally, ECC is employed to encrypt device requests, ensuring that sensitive information is protected during transmission.

\subsubsection{Integrity}

The integrity of cross-domain data and communications is maintained through the use of secure encryption protocols and the dynamic weight adjustment mechanism in DFL. By continuously adjusting model weights based on real-time feedback, the system ensures that models accurately reflect the current state of device data, preserving the integrity of predictions and authorization decisions. Moreover, model compression techniques are employed to reduce data transfer, further ensuring that only essential information is shared, thus maintaining the integrity of transmitted data.

\subsubsection{Availability}

The availability of the proposed system is enhanced by the use of DFL and dynamic weight adjustment. DFL allows for continuous learning and adaptation, enabling the system to update authorization policies in response to changing environments and threats. The dynamic weight adjustment mechanism ensures that models are optimized for the specific characteristics of each domain, improving the accuracy of authorization decisions and ensuring that legitimate devices have access to resources. This approach enhances the system's resilience to attacks and failures, ensuring that services remain available to authorized devices.

\subsubsection{Access Control}

The proposed security mechanism involves issuing one-time tokens after pre-authorization decisions are made. Once a device from $Dom_{A}$ is deemed trustworthy, it receives a token that allows it to access specified resources in $Dom_{B}$ directly. This token-based approach simplifies the access control process, reducing the overhead associated with repeated authentication and authorization procedures.

Tokens are generated with high entropy to ensure uniqueness and are bound to specific access privileges, which can be dynamically adjusted based on real-time data. This granular control over access permissions helps mitigate the risk of unauthorized access, as tokens are tied to the device's identity, role, and contextual factors at the time of issuance.

\subsection{Performance Evaluation}

\subsubsection{Experiment Settings}

All experiments were conducted in a Windows environment on an Intel(R) Core(TM) i7-13700KF, 3400 Mhz CPU, NVIDIA GeForce RTX 4080, and 32 GB of RAM. The programming language used for the implementation was Python, and the DFL framework employed was PyTorch.

We simulated a multi-domain network environment, where each network domain consists of a set of mobile devices and edge nodes. In each network domain, we simulated a zero-trust authentication and authorization system, including AM, authorization modules (PEP, PE, TE and data storage system), and CAM, to manage access requests to resources. The AM is responsible for device registration and authentication, the authorization module makes authorization decisions based on device information and access request attributes, and the CAM collects device context information to enhance the efficiency and security of cross-domain authentication and authorization. 

In each network domain, we modeled various access request scenarios to test the system’s performance. The simulations included diverse device behaviors and access patterns to represent realistic usage conditions. To ensure a realistic simulation environment, we incorporated a network latency of 10ms for cross-domain transmissions.

The DFL framework was implemented using PyTorch and configured with the following hyperparameters: a base learning rate of 0.01, a batch size of 32, and 100 training epochs. Each domain trained a local model on its dataset, sharing only model parameters with other domains. A dynamic weight adjustment mechanism based on real-time feedback was employed to ensure accuracy and robustness.

\subsubsection{Computation Overhead}

We evaluate the computational overhead intra and cross domains by theoretically analyzing each step of the authentication and authorization process. Here, $Exp$ denotes exponentiation, $H$ is a typical hash function, $Sig$ is a single signing operation using a private key, $I$ is the cost of performing one model inference, $CP$ is the cost of executing one policy judgment and decision, $M$ represents scalar multiplication, and $CS$ is the cost of performing one symmetric operation (encryption or decryption using symmetric keys).

Table I provides the computational overhead incurred by devices during intra-domain and cross-domain authentication and authorization phases. The table shows that intra-domain authentication and authorization require fewer computational resources compared to cross-domain processes. The primary reason is the additional costs associated with encrypting cross-domain transmissions and generating one-time tokens during pre-authorization. The total computational overhead for intra-domain operations is calculated as $3Exp+2H+Sig+2I+CP$, while for cross-domain operations it is $3Exp+3H+2Sig+4M+2CS+2I+CP$.

This analysis demonstrates that while cross-domain operations introduce additional computational overhead, they are necessary to ensure the security and integrity of cross-domain communications and data sharing. The additional costs are primarily due to the encryption and token generation processes, which are critical for maintaining a secure zero trust environment.

\begin{table*}[htbp]
    \centering
    \captionsetup{justification=centering, labelsep=newline}
    \caption{Computation Overhead of Each Step\label{tab:table1}}
    \begin{tabular}{|c|c|c|c|c|>{\centering\arraybackslash}m{2.5cm}|}
    \hline
    Step & Registration & Cross-domain Transmission & Authentication & Authorization & Total \\
    \hline
    Intra-Domain & $2Exp+H+Sig$ & $/$ & $Exp+H$ & $2I+CP$ & $3Exp+2H+Sig+2I+CP$ \\
    \hline
    Cross-Domain & $2Exp+H+Sig$ & $4M+2CS$ & $Exp+H$ & $2I+CP+H+Sig$ & $3Exp+3H+2Sig+4M+2CS+2I+CP$ \\
    \hline
    \end{tabular}
\end{table*}

\subsubsection{Latency}

\begin{figure*}[!t]
\centering
% 子图 (a)
\begin{subfigure}[b]{0.48\textwidth}
    \centering
    \includegraphics[width=\textwidth]{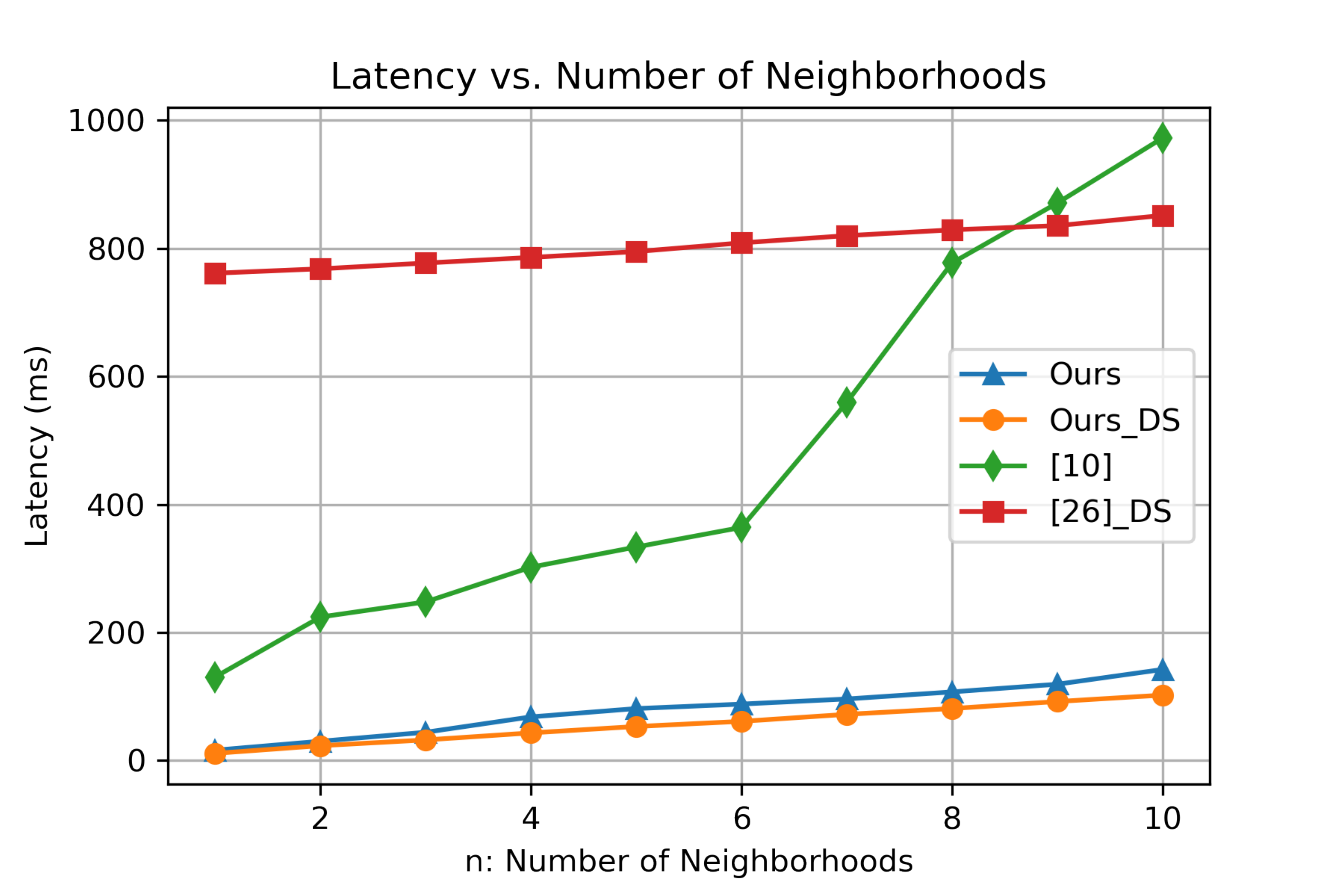}
    \caption{Neighboring domains $n$.}
    \label{fig_5a}
\end{subfigure}
\hfill
% 子图 (b)
\begin{subfigure}[b]{0.48\textwidth}
    \centering
    \includegraphics[width=\textwidth]{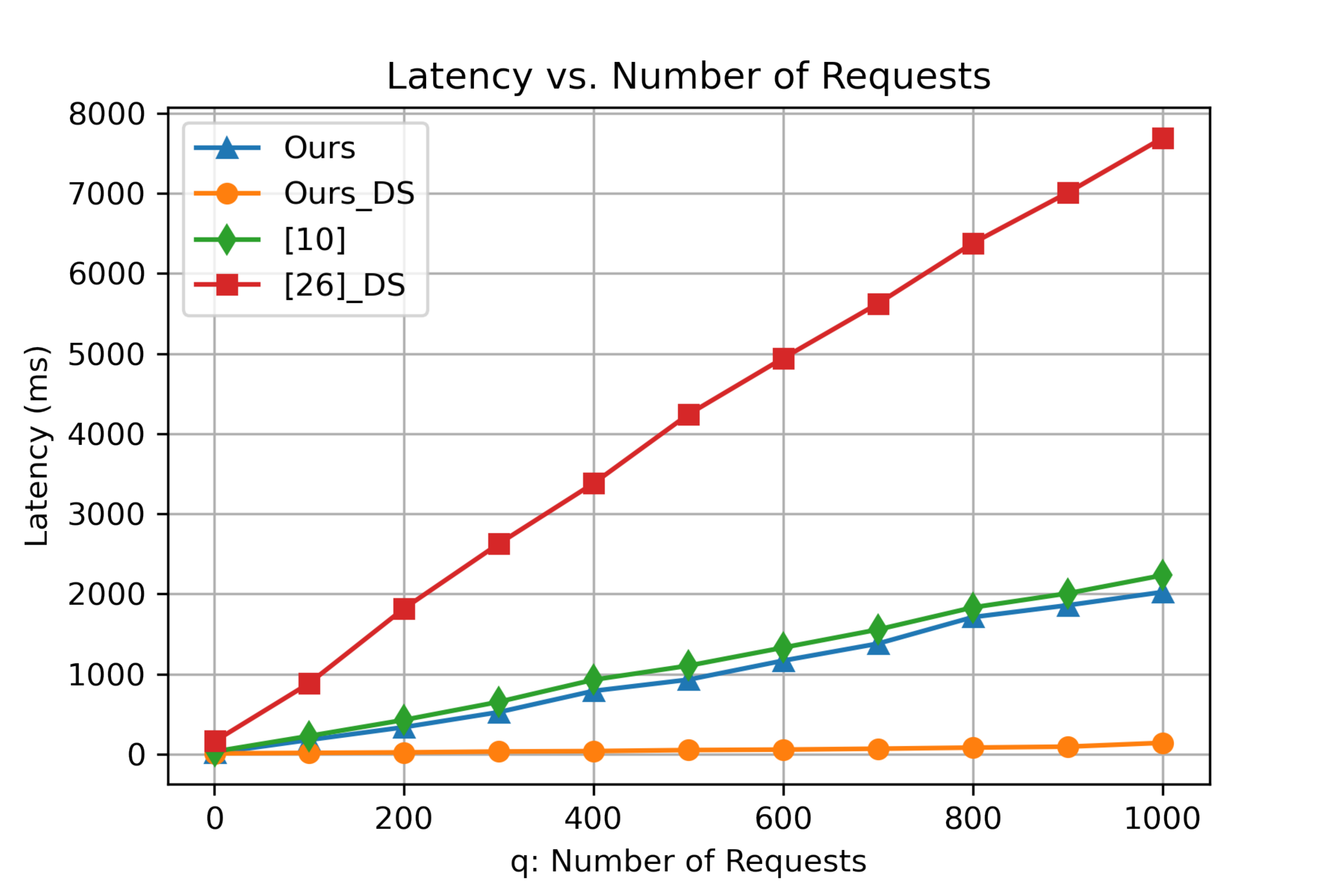}
    \caption{Requests $q$.}
    \label{fig_5b}
\end{subfigure}

% 主标题
\caption{The latency with varying (a) neighboring domains $n$ and (b) requests $q$.}
\label{fig_5}
\end{figure*}

We compared the proposed scheme with cross-domain authentication scheme \cite{feng2021blockchain} and the cross-domain data sharing scheme \cite{liu2023secure}. Since the scheme in \cite{liu2023secure} is a cross-domain data sharing protocol, we experimented not only with the latency of the entire pre-authorization process (i.e., from the device issuing a request across domains to receiving a one-time token) but also with the latency of the data sharing process (i.e., the request being transmitted from the source domain to the target domain). We set a target domain, where $n$ represents the number of neighboring domains and $q$ represents the number of parallel requests. Fig.~\ref{fig_5} illustrates how the latency of the authorization process varies with $n$ and $q$.

Fig.~\ref{fig_5a} illustrates that as the number of neighboring domains increases, the latency also increases to varying degrees. Our scheme exhibits significantly lower latencies compared to schemes in \cite{feng2021blockchain} and \cite{liu2023secure}, both in terms of the entire pre-authorization process and the data sharing part. This is because our scheme transmits only a small amount of data during the cross-domain authentication and authorization process, just enough to satisfy the information required for model inference in the target domain, thereby reducing network load and achieving lower latency. The model inference process in the target domain also incurs some latency, which determined by the complexity of the model. Fig.~\ref{fig_5b} demonstrates that as the number of parallel requests increases, the latency correspondingly increases. Although our scheme introduces some latency in the model inference stage, it still outperforms schemes \cite{feng2021blockchain} and \cite{liu2023secure}. Furthermore, these latencies occur only during the pre-authorization process. Once devices are authorized, they can directly utilize one-time tokens for access across domains, with token authentication times negligible, thus enabling real-time cross-domain access for mobile devices.

\subsubsection{Throughpput}

\begin{figure}[t]
  \centering
  \includegraphics[width=1.0\linewidth]{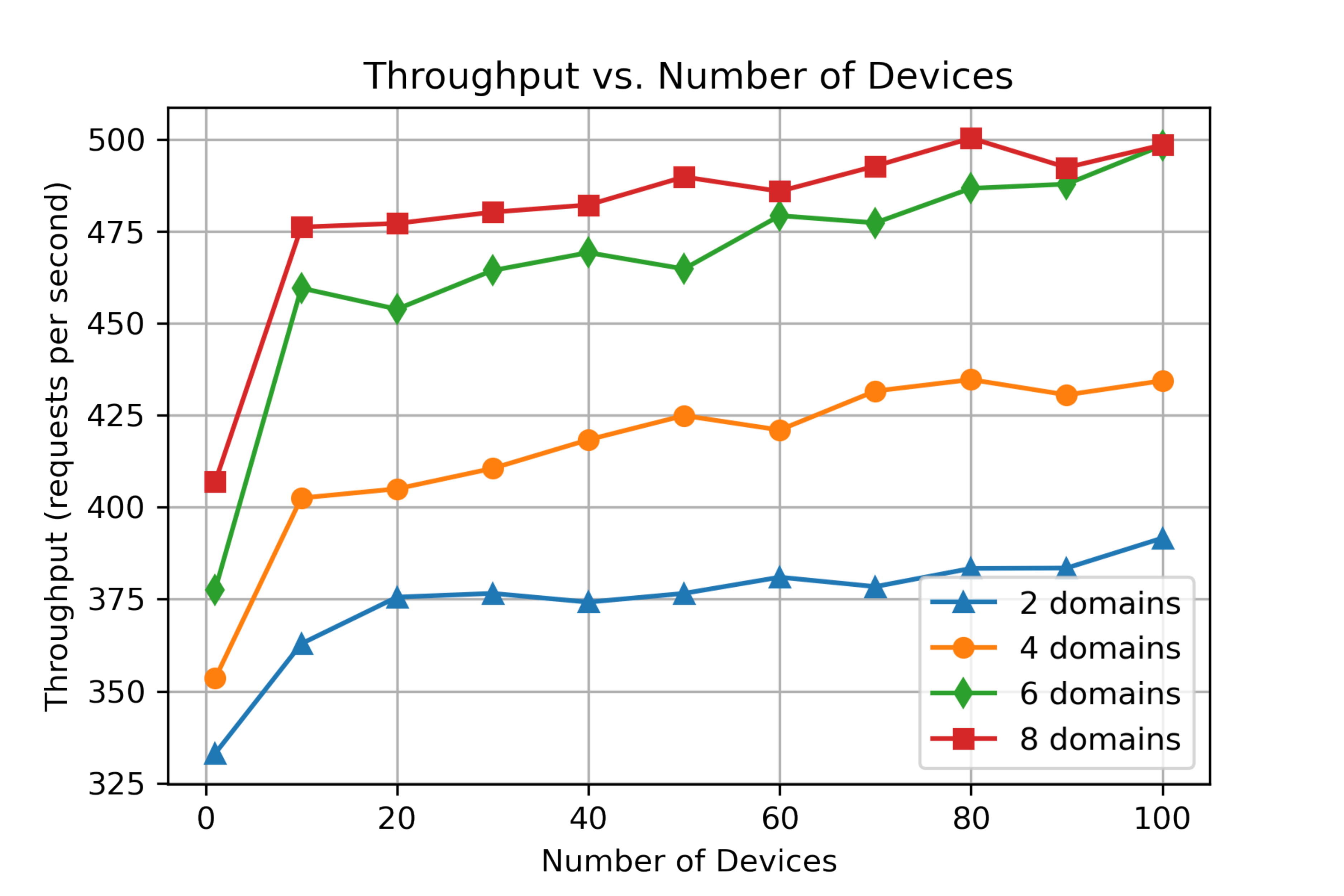}
  \caption{The throughput with varing devices number.}
  \label{fig_6}
\end{figure}

In Fig.~\ref{fig_6}, we present the throughput variation with the change in the number of devices across different numbers of neighboring domains. Each round of experiments involved 1000 requests, evenly distributed among the devices in each round. It can be observed that as the number of devices increases, the throughput initially rises correspondingly. However, when the number of devices reaches 10-20 and beyond, the system concurrently handles too many requests, leading to resource contention issues. This subsequently affects performance and restricts further growth in throughput. Consequently, the increase in throughput becomes minimal in the later stages, stabilizing at a relatively constant level.

Meanwhile, with the increase in the number of domains, the throughput also correspondingly increases. From Fig.~\ref{fig_6}, it can be observed that when the number of devices reaches 100, the throughput for only two domains is 390 r/s, while in cases with 6 and 8 domains, the throughput can reach 500 r/s. This is because increasing the number of domains enhances the system's ability to concurrently handle requests. As the number of domains increases, the system can process more requests simultaneously, avoiding overloading of individual nodes and effectively utilizing system resources, thereby improving the overall throughput.

\section{Conclusion}
In this paper, we have proposed a novel framework that integrates DFL with ZTA to enhance cross-domain authentication and authorization in IoT networks. The proposed scheme significantly improves security, efficiency and data privacy. By combining DFL, we ensure that sensitive data remains decentralized, thereby enhancing privacy. The analysis and simulation experiments show that the incorporation of ZTA improves security by enforcing strict access controls and continuous verification, while the proposed optimized protocols significantly reduce computational overhead and latency, enhancing overall efficiency.

The proposed DFL-based ZTA framework presents a significant advancement in securing cross-domain interactions within IoT networks, laying a strong foundation for future IoT innovations and applications. The diverse potential applications of the proposed framework span smart cities, healthcare, industrial IoT, and supply chain management, where the need for secure and efficient cross-domain interactions is crucial. Future work will aim to refine federated learning algorithms, enhance real-time feedback mechanisms, and validate the framework through practical real-world implementations.

% or


% ============================================
%\appendices
%Appendix one text goes here %\cite{Roberg2010}.

%\section{}

%\section*{Acknowledgment}

%The authors would like to thank D. Root for the loan of the SWAP. The SWAP that can ONLY be usefull in Boulder...

\ifCLASSOPTIONcaptionsoff
  \newpage
\fi



%\begin{thebibliography}{1}


\bibliographystyle{IEEEtran}
\bibliography{main}
%\end{thebibliography}
%\begin{biography}[{\includegraphics[width=1in,height=1.25in,clip,keepaspectratio]{mshell}}]{Michael Shell}

% ==== SWITCH OFF the BIO for submission
% ==== SWITCH OFF the BIO for submission

%% insert where needed to balance the two columns on the last page with
%%\newpage

% ==== SWITCH OFF the BIO for submission
% ==== SWITCH OFF the BIO for submission


\vfill


\end{document}